\documentclass{article}
\usepackage{spconf,amsmath,graphicx}

\title{Unsupervised Data-Driven Nuclei Segmentation for Histology Images}

\name{Vasileios Magoulianitis, Peida Han, Yijing Yang, C.-C. Jay Kuo}

\address{University of Southern California, Los Angeles, California, USA}

\begin{document}
%
\maketitle
\begin{abstract}

An unsupervised data-driven nuclei segmentation method for histology
images, called CBM, is proposed in this work. CBM consists of three
modules applied in a block-wise manner: 1) data-driven color transform
for energy compaction and dimension reduction, 2) data-driven
binarization, and 3) incorporation of geometric priors with
morphological processing. CBM comes from the first letter of the three
modules -- ``Color transform", ``Binarization" and ``Morphological
processing".  Experiments on the MoNuSeg dataset validate the
effectiveness of the proposed CBM method. CBM outperforms all other
unsupervised methods and offers a competitive standing among supervised
models based on the Aggregated Jaccard Index (AJI) metric. 

\end{abstract}

\section{Introduction}\label{sec:intro}  

Histology images provide strong cues to pathologists in cancer diagnosis
and treatment. Automated nuclei instance-level segmentation for
histology images provides not only the number, density and sizes of
nuclei but also morphological features, such as the magnitude and the
cytoplasmic ratio. This information facilitates cancer diagnosis and
assessment of the tumor aggressiveness rate.  Nuclei segmentation tasks
can be conducted in a supervised or an unsupervised manner.  For
supervised methods, the annotation of high resolution histology images
in pixel-level accuracy is a time-consuming job, being carried out by
expertized physicians. Other identified challenges include the
variability of cell appearance from different organs, unclear
boundaries, as well as color and intensity variations in stained images from different laboratories. Furthermore, it is a subjective task
and annotated labels tend to vary from one person to the other. All the
above factors challenge the practical generalizability of supervised
segmentation methods, as histology images become larger in their number
and more diversified in content. 

Earlier methods on nuclei segmentation were mainly unsupervised.  They
were based on thresholding \cite{naik2008automated, win2017automated},
mathematical morphology for robust feature extraction
\cite{thiran1996morphological}, or statistical modeling for segmentation
and likelihood maximization for boundary refinement
\cite{mouroutis1998robust}. Another popular tool was the watershed
algorithm, which was combined with various ways for extracting potential
nuclei locations \cite{malpica1997applying, veta2011marker}. Moreover, a work by Ali {\em et al.} \cite{ali2011adaptive} proposed an adaptive active
contour mechanism that takes boundary- and region-based energy terms of the cell into account. 

Recently, deep-learning-based (DL) methods \cite{zhou2019cia,
kumar2017dataset, chen2017dcan, liu2018path} have been applied to this
problem. A wide variety of DL models are trained by human labeled data.
In general, supervised methods provide significantly better performance
than the unsupervised ones. The motivation of our research stems from the
fact that nuclei instance labeling is a fairly laborious task, with
reportedly highly subjective annotations and miss-labeling rate
\cite{irshad2014crowdsourcing} that challenges the supervised solutions.  Several studies have been
made to mitigate the labeling cost.  For example, Qu {\em et al.}
\cite{qu2020weakly} proposed a two-stage learning framework using coarse
labels.  Other researchers \cite{sahasrabudhe2020self, xie2020instance}
investigated the self-supervised DL method to reduce the number of
required labeled data by exploiting the observation that nuclei size
and texture can determine the magnification scale. Besides, there is a domain shift problem \cite{yan2019domain} arising
from stain and nuclei variations in histology images of different
organs, patients and acquisition protocols.

\begin{figure}[t]
\begin{center}
\includegraphics[width=.4\linewidth]{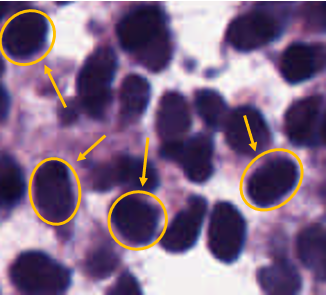}
\end{center}
\caption{Illustration of nuclei cell appearance in histology images.}
\label{fig:example}
\end{figure}

\begin{figure*}[t]
\begin{center}
\includegraphics[width=1\linewidth]{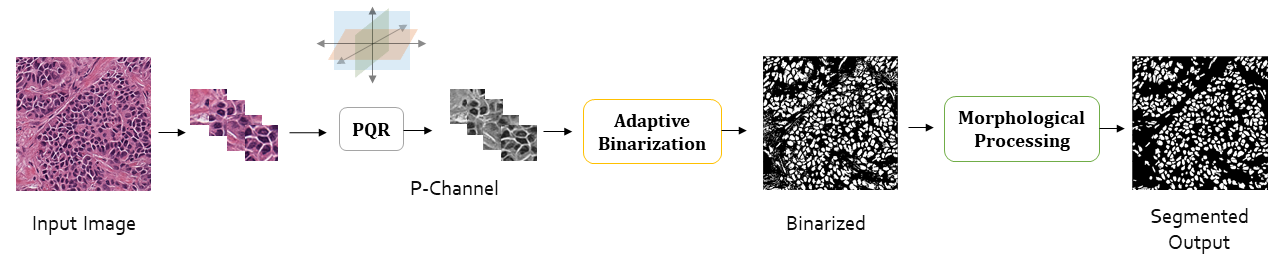}
\end{center}
\vspace{-5mm}
\caption{An overview of the proposed CBM method.}
\label{fig:Pipeline}
\end{figure*}

In spite of the high performance, DL-based solutions have their own
shortcomings on top of the high labeling cost.  First, they are
perceived as a ``black-box" approach. Interpretable solutions are highly
desired in medical applications as they enable explainable decisions and
the tools are more trustworthy in clinical use. Second, they bear a high
computational cost in training and testing \cite{LAL2021104075} because
of a large number of parameters in order to achieve a certain
performance. 

In this work, we propose an unsupervised data-driven method to solve the
nuclei segmentation problem. The solution consists
of three modules applied to each image block of size $50\times 50$: 1)
data-driven color transform for energy compaction and dimension
reduction, 2) data-driven binarization, and 3) incorporation of
geometric priors with morphological image processing.  It is named
``CBM" because of the first letter of the three modules -- ``Color
transform", ``Binarization" and ``Morphological processing". 

We conduct experiments on the MoNuSeg dataset to demonstrate the
effectiveness of the proposed CBM method.  It outperforms all other
unsupervised approaches and stands in a competitive position with
supervised ones based on the Aggregated Jaccard Index (AJI) performance
metric.  Thus, the proposed CBM method is an attractive solution in
practice nuclei segmentation, since it yields comparable performance with
state-of-the-art supervised methods while requiring no training labels. 
 
\section{Proposed CBM Method}\label{sec:Meth} 

Each histology test image of the experimenting dataset  has a spatial resolution of $1000 \times 1000$
pixels and each pixel has R, G and B three color channels. We partition
each image into nonoverlapping blocks of size $50 \times 50$ pixels and
apply contrast enhancement to accentuate the boundaries around nuclei as
a pre-processing step. Then, CBM processes each block independently with
three modules as elaborated below.

\vspace{-1ex}
\subsection{Data-Driven Color Transform}\label{subsec:color}
\vspace{-1ex}

\begin{figure}[ht]
\centering
\includegraphics[width=0.2\textwidth]{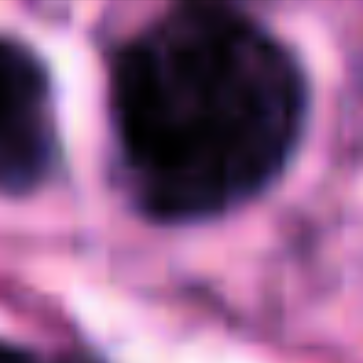} 
\includegraphics[width=0.2\textwidth]{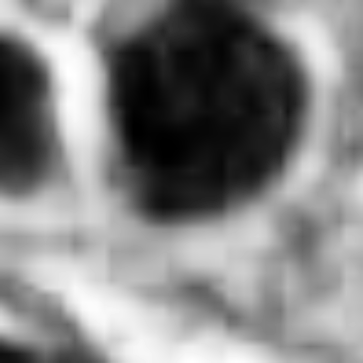} \\
(a) \hspace{19ex} (b) \\
\vspace{2mm}
\hspace{-1mm}
\includegraphics[width=0.35\textwidth]{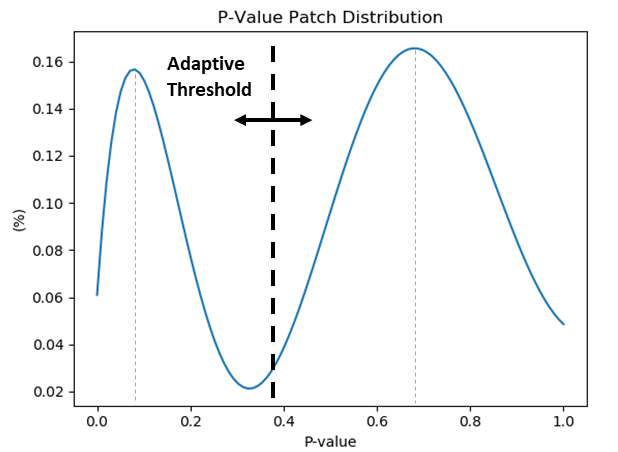} \\
\hspace{1ex} (c)
\caption{Comparison of two representations in a block image: (a) R/G/B
color and (b) $P$ value in gray and (c) its corresponding bi-modal 
histogram.}\label{fig:example_2} \end{figure}

Nuclei segmentation is a binary decision problem for each pixel; namely,
it belongs to either the nucleus or the background region. We need pixel
attributes to make decision. A pixel has R/G/B color values in raw
histology images. There exist strong correlations between RGB channels
in histology images as shown in Fig. \ref{fig:example_2}(a). We can
exploit this property for energy compaction. 

There are many well-known color spaces such as YUV, LAB, HVS, etc.
However, they are not data-dependent color transforms. To achieve
optimal energy compaction, we apply the principal component analysis
(PCA) to the RGB 3D color vector of pixels inside one block. That is, we
can determine the covariance matrix between the R/G/B color components
based on pixels in the region. Then, the three eigenvectors define the
optimal transform and their associated eigenvalues indicate the energy
distributions among the three channels. The transform output channels
are named $P$, $Q$ and $R$ channels. They are channels of the first, the
second and the last components, respectively. 

The energy distributions of three color components of the RGB, LAB and
PQR color spaces for a representative histology image is shown in Table
\ref{tab:color}, where the P/Q/R channel energy distribution is averaged
over all blocks in one image. As shown in the table, the first principal
component, $P$, has an extremely high energy percentage while the rest
two components have very limited energy. As a result, we can treat the
latter two as background noise and discard them.  Instead of considering
segmentation using three attributes, we simplify it greatly using a
single attribute. As the first principal subspace, $P$ points at the
direction where the variance is maximized. It better captures the
transition from background to nuclei cell areas, thus leading to a more
distinct local histogram. 

We normalize raw $P$-channel values to the range of $[0,1]$ with linear
scaling.  Note that, if ${\bf x}$ is an eigenvector of the covariance
matrix of RGB color channels, $-{\bf x}$ is also an eigenvector. We
choose the one that maps a higher $P$ value to background (i.e.,
brighter) and a lower $P$ value to nuclei (i.e., darker). This can be
easily achieved by imposing the $P$ value to be consistent with the
luminance value in the LAB color space, since the background luminance
is higher than those of nuclei. The original color and the normalized
P-channel representations for a block are compared in Fig.
\ref{fig:example_2}.  Performance using the $P$ channel against the $L$
channel (in the LAB color space) will be compared in Sec.
\ref{sec:Results}. 

\begin{table}[h]
\caption{Energy distribution of three channels of three color 
spaces in a representative histology image.}\label{tab:color}
\begin{center}
\begin{tabular}{|c|c|c|c|}\hline
RGB        & LAB      & PQR      \\ \hline
38.1\% (R)   & 29.3\% (L) & 97.7\% (P) \\ \hline
27.2\% (G)   & 40.1\% (A) &  1.9\% (Q) \\ \hline
34.7\% (B)   & 30.6\% (B) &  0.4\% (R) \\ \hline
\end{tabular}
\end{center}
\end{table}

\vspace{-1ex}

One can see that the energy compaction capability of P/Q/R depends on
color homogenuity in a region. In most images, as the block size grows
larger, the $P$-channel energy compaction property becomes lower. This
is attributed to the fact that the color distribution of a larger block
is less homogeneous and the correlation structure is more complicated. 

\vspace{-2ex}
\subsection{Data-Driven Binarization} \label{subsec:clustering}
\vspace{-1ex}

To conduct the binary classification of pixels, we study the histogram
of the $P$ value of each pixel in a block. A representative histogram is
shown in Fig. \ref{fig:example_2}(c), which has a bi-modal distribution.
The modality in the left is contributed by the $P$ values of pixels in
the nuclei region while that in the right comes from the $P$ values of
pixels in the background region. There exists a valley between the two
peaks, which are from pixels lying on transitional boundaries between
nuclei and background. Thus, binarization can be achieved by identifying
the intermediate point in between the two modalities and use the
associated $P$ value as the adaptive threshold, which is denoted by $T$.
A pixel of $P$ value is classified to the background region if $P > T$
and the nuclei region if $P \leq T$. Since threshold $T$ is determined
by the histogram of the $P$ value of pixels in a block, the binarization
process is fundamentally an adaptive thresholding method. Whether it
will yield a successful outcome depends on the bi-modal histogram assumption. 

The bi-modal histogram assumption holds under the following two conditions:
\begin{enumerate}

\item if the block size is not too large,
\item if the ratio of the nuclei pixel number and the background pixel 
number does not deviate much from unity. 
\end{enumerate}
In case the first condition is not met, we may see $K$-modalities with
$K > 2$.  Then, there are multiple valley points, which makes the
threshold selection challenging. If the second condition is not met, it
means that the majority of pixels belong to one of two classes and we
may see one dominant modality while the second modality is weak. Because
it is difficult to choose a robust threshold under the first condition,
we partition one block of size $50 \times 50$ into four sub-blocks of
size $25 \times 25$ and conduct the data-driven color transform and
binarization in each of the four sub-blocks. On the other hand, if the
second condition happens, we merge 4 blocks into a super-block of size
$100 \times 100$ and conduct the same processing in the super-block. 

\vspace{-1ex}
\subsection{Morphological Processing}\label{subsec:geometry}
\vspace{-1ex}

We have so far concentrated on pixel classification based on its color
attributes in the first two modules of the CBM method. Now, we would
like to take the neighborhood of a pixel into account. Typically, nuclei
appear in form of rounded blobs (i.e., convex objects). Yet, we observe
the following three common errors after the second module:
\begin{itemize}
\itemsep -1ex
\item[(a)] nuclei instances may be falsely connected,
\item[(b)] false positives may appear because of the staining process,
\item[(c)] holes exist inside the nuclei cell because of the inner intensity variations. 
\end{itemize}

\begin{figure}[h]
\begin{center}
\includegraphics[width=.7\linewidth]{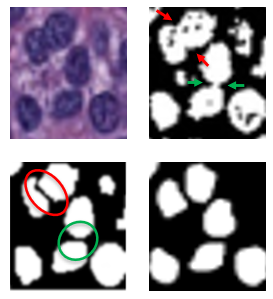}
\end{center}
\vspace{-3ex}
\caption{Visualization of the morphological processing effect: input
block (upper left), noisy binarized output (upper right), an improved
result by splitting two distinct nuclei (bottom left), and segmentation
ground-truth (bottom right).} \label{fig:PostProcGif}
\end{figure}

For (a), we can split the larger ones using the convex hull algorithm to
find high convexity areas that imply an underlying connectivity between
two cells.  For (b), we can filter out abnormally small nuclei in
the first place based on the prior knowledge on the average area of a
nucleus. For (c), the hole filling filter is used to correct the false
negatives inside the nuclei cell.  For some images with dense cell areas
and blurred boundaries, the connectivity may be more severe, having
bundled together more than two cells. We found that, after applying the
previous procedure in an iterative manner, the performance further
increases. Thus, we repeatedly remove small cells and split larger
pieces --with relatively high convexity areas--, until no further
changes in the full image.  The effect of this processing is illustrated
in an example as shown in Fig.  \ref{fig:PostProcGif}. 

\vspace{-1ex}
\vspace{-1ex}

\section{Experimental Results}\label{sec:Results}

The data from the 2018 MICCAI MoNuSeg Challenge \cite{kumar2017dataset}
are used to evaluate the performance of our proposed CBM method. This is
a popular dataset for existing work, providing also different testing
configurations. For the validation purpose, we follow the data
splitting as suggested in \cite{kumar2017dataset}. 

\begin{table}[]
\caption{Comparative results of different methods using the AJI on Test-$2$ set ($6$
histology images from bladder, colon, stomach) and the testing data ($14$ images) from the
MoNuSeg Challenge, where the best performance is shown in boldface.}\label{tab:CompResults}
\begin{center}
\begin{tabular}{|c|c|c|} \hline
Method  & \hspace{2mm} Test-$2$ \hspace{2mm} & Challenge Data     \\ \hline
\multicolumn{3}{|c|}{{\em Unsupervised}} \\ \hline
\rule[-1ex]{0pt}{3.5ex}  Cell Profiler \cite{carpenter2006cellprofiler}  & 0.0809 & 0.1232 \\ \hline
\rule[-1ex]{0pt}{3.5ex}  Fiji \cite{schindelin2012fiji} & 0.3030 & 0.2733 \\ \hline
\rule[-1ex]{0pt}{3.5ex}  Self-Supervised \cite{sahasrabudhe2020self} & - &  0.5354 \\ \hline
\rule[-1ex]{0pt}{3.5ex}  CBM (Ours) & \bf{0.5808} & \bf{0.6142}   \\ \hline
\multicolumn{3}{|c|}{{\em Supervised}} \\ \hline
\rule[-1ex]{0pt}{3.5ex}  CNN2 \cite{kumar2017dataset} & - & 0.3482  \\ \hline
\rule[-1ex]{0pt}{3.5ex}  CNN3 \cite{kumar2017dataset} &  0.4989 & 0.5083  \\ \hline
\rule[-1ex]{0pt}{3.5ex}  DCAN \cite{chen2017dcan} & 0.5449 & -  \\ \hline
\rule[-1ex]{0pt}{3.5ex}  PA-Net \cite{liu2018path} & 0.5608 & - \\ \hline
\rule[-1ex]{0pt}{3.5ex}  BES-Net \cite{oda2018besnet} & 0.5823 & -  \\ \hline
\rule[-1ex]{0pt}{3.5ex}  CIA-Net \cite{zhou2019cia} &  \bf{0.6306} & 0.6907 \\ \hline
\rule[-1ex]{0pt}{3.5ex}  SSL \cite{xie2020instance} &  - & \bf{0.7063}   \\ \hline
\end{tabular}
\end{center}
\end{table}

For performance benchmarking with other existing works, we report the
results on the 6 histology images set from bladder, colon,
stomach, referred as Test-2, as well as the test data from the challenge set (14 images from
various organs) in Table \ref{tab:CompResults}, where the commonly used
AJI metric \cite{kumar2017dataset} is adopted.  The AJI metric is more
accurate for instance-level segmentation tasks by taking both the nuclei
level detection and pixel-level error performance into account.  

As shown in the table, our method outperforms all other unsupervised
ones by a large margin in either testing scenarios. Notably, the gap is
large even for the DL-based unsupervised approach denoted by
``self-supervised" \cite{sahasrabudhe2020self} in the table. As compared
with supervised methods, CBM stands at a competitive position. It
achieves similar performance with sophisticated models such as BES-Net
\cite{oda2018besnet}.  It is also worthwhile to emphasize that most of
the supervised methods use expensive models as a backbone architecture (in terms
of the number of its parameters) such as Res-Net or Dense-Net.
Instead, CBM is a parameter-free method with simple components. Its
computational complexity is significantly lower. 

It is also evident from Table \ref{tab:CompResults} that supervision may
help to reach higher performance. Nevertheless, it is challenging for
most DL supervised methods to generalize well in this problem, especially
when the training data are in paucity.  For instance, the current
state-of-the-art SSL method achieves the reported performance using the
full training set. According to their weakly supervision analysis
\cite{xie2020instance}, SSL achieves comparable performance with CBM
when trained with roughly 50\% of the data. 
 
To illustrate the energy compaction capability of the PQR channel
decomposition, we compare the AJI performance for the $P$ channel and
the $L$ channel (in the LAB color space) in Table \ref{tab:Ablation}.
The advantage of using the $P$ channel is clearly demonstrated in the
table.

\vspace{-2ex}
\begin{table}[t]
\caption{AJI performance comparison between $P$ and $L$ channels.}
\label{tab:Ablation}
\begin{center}       
\begin{tabular}{|c|c|c|} 
\hline
\rule[-1ex]{0pt}{3.5ex}  Colorspace & \hspace{2mm} Test-$2$ \hspace{2mm} & Challenge Data  \\ \hline
\rule[-1ex]{0pt}{3.5ex}   L  & 0.5414 & 0.5856  \\ \hline
\rule[-1ex]{0pt}{3.5ex}   P  & {\bf 0.5808} & {\bf 0.6142}  \\ \hline
\end{tabular}
\end{center}
\end{table}

\section{Conclusion and Future Work}\label{sec:conclusion}
 
Nuclei segmentation in histology images is a demanding and prone to
errors task for physicians, and its automation is of high importance for
cancer assessment. The proposed CBM method offers a promising
lightweight unsupervised direction for nuclei segmentation, requiring no
labeled data and no parameter cost. It addressed the problem based on
data-driven color conversion and binarization, as well as a morphological module
that takes into account priors about the nuclei shape and size. In the future, we would
like to boost the segmentation performance furthermore by exploring
weakly or self supervision. 

\newpage

{\small
\bibliographystyle{IEEEtran}
\bibliography{strings,refs}
}
\end{document}